\documentclass[12pt,preprint]{aastex}

\usepackage{emulateapj5}
\usepackage{apjfonts}
\usepackage{epsf}

\slugcomment{Accepted for publication in ApJL}

\shorttitle{Searching for dark matter halos}
\shortauthors{Miyazaki et al.}

\begin{document}
\title{Searching for dark matter halos in the Suprime-Cam 2 sq deg
field}

\author{
Satoshi Miyazaki\altaffilmark{1}, 
Takashi Hamana\altaffilmark{1}, 
Kazuhiro Shimasaku\altaffilmark{2}, 
Hisanori Furusawa\altaffilmark{1},
Mamoru~Doi\altaffilmark{3}, 
Masaru Hamabe\altaffilmark{4}, 
Katsumi Imi\altaffilmark{5},
Masahiko Kimura\altaffilmark{6}, 
Yutaka Komiyama\altaffilmark{1}, 
Fumiaki Nakata\altaffilmark{1},
Norio Okada\altaffilmark{1}, 
Sadanori Okamura\altaffilmark{2},
Masami Ouchi\altaffilmark{2}, 
Maki Sekiguchi\altaffilmark{6},
Masafumi Yagi\altaffilmark{1} and
Naoki Yasuda\altaffilmark{1}
}

\altaffiltext{1}{National Astronomical Observatory of Japan, Mitaka, Tokyo 181-8588, Japan}
\altaffiltext{2}{Department of Astronomy, University of Tokyo, Bunkyo, Tokyo 113-0033, Japan}
\altaffiltext{3}{Institute of Astronomy, University of Tokyo, Mitaka, Tokyo 181-0015, Japan}
\altaffiltext{4}{Department of Mathematical and Physical Sciences,Japan Women's University, Bunkyo, Tokyo 112-8681, Japan}
\altaffiltext{5}{Communication Network Center (Tsu-den) , Mitsubishi Electric, Amagasaki, Hyogo 661-8661, Japan}
\altaffiltext{6}{Institute for Cosmic Ray Research, University of Tokyo, Kashiwa,Chiba 277-8582, Japan}

\begin{abstract}
We report the first result of weak gravitational lensing survey 
on a 2.1 deg$^{2}$ $R_c$-band image taken with a wide field camera 
(Suprime-Cam) on the prime focus of 8.2 m Subaru Telescope. 
The weak lensing mass reconstruction is applied to the data to search
for dark matter halos of cluster scale; $M \ge 10^{14} M_{\odot}$.
The reconstructed convergence field is divided by 1-$\sigma$ noise 
to obtain the signal-to-noise ratio map (S/N-map) of the detection.
Local maxima and minima are searched on the S/N-map and the 
probability distribution function (PDF) of the peaks are created 
to compare with model predictions. 
We found excess over noise PDF created from the randomized realization 
on both positive and negative sides.
Negative peaks imply the presence of voids in the dark matter 
distribution and this is the first report of the detection. 
Positive peaks, on the other hand, represent the dark matter 
halos and the number count  of the halos on the 2.1 deg$^2$ 
image is $4.9 \pm 2.3$ 
for S/N $>$ 5 where the Gaussian smoothing radius of the 
convergence map is $1'$. 
The result is consistent with the prediction assuming the
Press-Schechter mass function and the NFW halo profile under 
the cluster normalized CDM cosmology. 
Although the present statistics is not enough due to the limited field
of view, this work demonstrates that dark matter halo count via weak 
lensing is a promising way to test the paradigm of 
structure formation and cosmological model.
\end{abstract}


\section{Introduction} 

Dark matter halos, which are the virialized systems formed via the 
gravitational amplification of initial density fluctuations, are 
one of the most important cosmological probes. 
Since their evolution is governed purely by gravity,
their number density and the evolution of clustering properties 
can be precisely predicted analytically as well as numerically.
Weak lensing is a unique observational technique to detect dark matter 
halos because it detects halos not by their {\it light} but by more 
essential physical parameter, the {\it mass}. It also allows 
us to directly estimate the halo mass bypassing a use of ambiguous 
empirical relations like the mass-temperature (or velocity dispersion) 
or mass-luminosity relations. Therefore, comparison of theories and 
weak lensing observations is a direct and reliable way to test 
the paradigm of the structure formation and to constrain the 
cosmological parameters.

Theoretical prescriptions for predicting statistics of weak lensing
halos have been developed so far.
\cite{kruseschneider99} first calculated the expected number of 
dark matter halos detected by means of the aperture 
mass method \citep{schneider96}.
Predicted number density of halos
exceeds 10 per deg$^2$ for cluster-normalized cosmologies if one
obtains moderately deep images whose galaxy density is $n_g \sim$30 
arcmin$^{-2}$. \cite{bartelmanetal01} pointed out that the number
density is sensitive to the density profile of halos.
On the other hand, \cite{jainvanwaerbeke00} discussed the statistics 
of the {\it peaks} on convergence map.
They claimed that the statistics can be an alternative approach to 
probe the mass function of dark matter halos and to constrain 
$\Omega_m$.

Observationally, weak lensing has become reliable measures to map
the dark matter distribution in clusters of galaxies 
(see a review by Mellier 1999). Recently, serendipitous 
detections of mass concentrations outside clusters on a 
reconstructed mass map have been reported by 
several authors 
\citep{erbenetal00,umetsufutamase00,wittmanetal01,mirallesetal02}.
All cases have no clear optical counterparts except the one 
by \cite{wittmanetal01} which is identified as a cluster of galaxies 
at $z = 0.276$. 

These findings demonstrate that dark matter halo search via weak
lensing is feasible and can detect even lower luminosity dark matter
halos that are overlooked in conventional optical cluster surveys.
The number of samples obtained  so far is, however, limited. 
In order to make statistically significant discussions, weak lensing 
survey over a representative cosmic volume is crucial.
The survey requires not only wide field  but also high image
quality to make reliable lensing analysis. Suprime-Cam on the
8 m Subaru telescope is the most ideal camera for the purpose.
The median seeing in $I_c$ band monitored over a year and a half 
is reported to be $\sim$ 0.6 arcsec \citep{miyazakietal02}.
In this {\it Letter}, we report the first results from the 
Suprime-Cam weak lensing survey that covers contiguous 2.1 
deg$^{2}$ field.
We examine statistical properties of local peaks on the reconstructed 
convergence map by comparing with model predictions.

\section{Observation and data reduction}

We set field size at 2.1 deg$^2$, which is the largest size 
possible to be accomplished during the commissioning 
phase of the Suprime-Cam. The field we chose is 
centered at R.A. = 16$^h$04$^m$43$^s$, 
decl. = +43$^{\circ}$12'19" (J2000.0). 
In order to compare light and mass we 
observed one of the seven fields that \cite{gho} made cluster 
surveys (GHO survey). The 2.1 deg$^2$ field has twelve GHO clusters 
and one Abell cluster (A 2158). Note, however, that the 
number density of clusters in this field is comparable to the one 
averaged over the seven GHO fields. 

We obtained  $R_c$-band images on the nights of 2001 April 23-25. 
Suprime-Cam provides a field of view of 34'$\times$ 27' with 
a scale 0".202 pixel$^{-1}$. 
Nine contiguous fields were observed in 3$\times$3 mosaic configuration. 
Each exposure on a given field was offset by 1' from other exposures 
to remove cosmic rays and defects on the CCDs. Total exposure time 
on each field was 1800 sec (360 sec $\times$ 5).

The individual images were de-biased and then flattened using a median
of all object frames taken during the observing run. 
Registration on
each field was first performed assuming a simple geometrical model. 
The parameters of the model include (1) optical distortion,  
(2) displacement and rotation of each CCD from the fiducial position and 
(3) the offset of pointing between the exposures \citep{miyazakietal02}. 
These parameters can be solved by minimizing position errors of
control stars common to all exposures and used for the first 
transformation of the images to a common coordinate system. After the 
transformation, the residual of the position of the control stars
is $\sim$ 0.5 pixel rms. The residual is due to several effects not
considered in the simple model including atmospheric dispersion and 
asymmetric aberration of optics. We note, however, that the residual 
vector is smoothly changing over position and can be well modeled as the 
third order bi-linear polynomial function of position. This model then 
gives the fine correction of the transformation function. As a result
the final residual decreased typically down to 0.07 pixel rms 
(14 miliarcsec). The seeing in the resulting image is 0".68 FWHM 
and the scatter among the fields is quite small of 0".04 rms.

We used the software suite {\it imcat}, an actual implementation of 
\cite{ksb95}, to carry out object detection, photometry and shape 
measurements of objects. Catalogs created for the nine fields are 
registered using stars on overlapping regions to result in a final 
catalog whose total field of view is 
1.64$^{\circ}$ $\times$ 1.28$^{\circ}$. 
Differences in photometric zero point among the fields due to 
variation of sky condition were compensated at this stage but the 
adjustment was not significant ($\sim$ 0.05 mag). For weak lensing
analysis, we adopt galaxies of $ 22 < R < 25.5 $ 
whose signal-to-noise (S/N) ratio, $nu$, calculated in the {\rm
imcat} exceed 15 \citep{baconetal01}. The number of objects in 
the catalog is 297547 (39.3 arcmin$^{-2}$) and the count peaks at R = 25.2.

\section{Galaxy shape analysis}
The estimate of the shear, $\vec{\gamma}$, from the observed
ellipticities of galaxies, 
$\vec{e} = \{I_{11} - I_{22}, 2I_{12}\} / (I_{11}+I_{22})$ involves
two steps where $I_{ij}$ are quadrupole moments of the surface brightness
of objects.
First, PSF anisotropy is corrected using the image of stars as
references, 
\begin{equation}
\label{Psmcorrection}
\vec{e'} = \vec{e} - \frac{P_{sm}}{P_{sm}^{*}}\vec{e^{*}},
\end{equation}
where $P_{sm}$ is the smear polarisability tensors and is
mostly diagonal \citep{ksb95}. 
$(P_{sm}^{*})^{-1}\vec{e^{*}}$ was calculated for stars scattered over
the field of view and we made fifth order bi-polynomial fit
to the values  as a function of the positions. 
This function is used in Eqn.(~\ref{Psmcorrection}) to correct the 
ellipticities of faint galaxies. The nine fields are treated 
independently in this procedure.
The rms value of ellipticities of the reference stars, $\langle
|\vec{e^{*}}|^2 \rangle ^{\frac{1}{2}}$, is reduced from 2.7\% to
1.3\% as a result of the correction. Note that the rms before 
the correction is already small thanks to the superb image quality of 
Subaru telescope.

\cite{lk97} worked out how to correct the ellipticities for 
the effect of the seeing. 
The {\it pre-}seeing shear $\vec{\gamma}$ is described as 
\begin{equation}
\vec{\gamma} = P_{\gamma}^{-1} \vec{e'}, \;\;
P_{\gamma} = P_{sh} - \frac{P_{sh}^{*}}{P_{sm}^{*}}P_{sm},
\end{equation}
where $P_{sh}$ is the shear polarisability tensor. The $P_{\gamma}$ of
individual galaxies are, however, known to be noisy estimate and 
we thus adopted smoothing and weighting method developed by
van Waerbeke et al.(2000; see also Erben et al. 2001 for a detail 
study of the smoothing scheme). 
For each object, twenty neighbors are first identified in the 
$r_g$-$magnitude$ plane where $r_g$ is a measure of size of 
objects which the {\it imcat} yields. 
A median value of $P_{\gamma}$ among the neighbors is adopted for a 
smoothed $P_{\gamma}$ of the object. 
The variance of raw $\gamma$ before the smoothing among the 
neighbors, $\sigma_\gamma^2$, is used to estimate the 
weight of the object, $w_n$, as
$w_n = 1/(\sigma_\gamma^2 + \alpha^2)$ where $\alpha^2$ is the variance 
of all the objects in the catalog and $\alpha\simeq 0.4$ here. 
Under the weighting scheme, an averaged value of a certain observable, 
$\langle Q \rangle$, is calculated as $\sum^N_{i=1}w_{n,i}Q_i / 
\sum^N_{i=1}w_{n,i}$ instead of $\sum^N_{i=1}Q_i/N$. 
The $\langle {\rm tr}(P_{\gamma}) \rangle$ for all objects is calculated as
$0.31$ but $P_{\gamma}$ depends on $r_g$.  For objects 
with $r_g > 3.5$, $\langle {\rm tr}(P_{\gamma}) \rangle$ is almost
constant, $\sim 0.65$, and becomes small below the $r_g$ down to 
$\langle {\rm tr}(P_{\gamma}) \rangle \sim 0.2$ for $r_g =
1.5$. 

\section{Results and discussions}
The convergence field, $\kappa$, is estimated using the
original \cite{kaiserandsquires93} inversion algorithm with 
a Gaussian window function 
$W(\theta)=\exp(-|\mbox{\boldmath{$\theta$}}|^2/\theta_G^2)/\pi\theta_G^2$. 
In order to estimate the noise of $\kappa$ field, we randomized
the orientations of the galaxies in the catalog and created 
a $\kappa_{\rm noise}$ map. The variance of the $\kappa_{\rm noise}$ 
depends on the smoothing scale 
$\theta_G$ and is represented as 
$\sigma_{\rm noise} \sim 0.023 \theta_G^{-1}$. This turned out to be
consistent with theoretical estimate, 
\begin{equation}
\sigma_{\rm noise} = \frac{\sigma_{\epsilon}}{2 \theta_G\sqrt{\pi n_g}},
\end{equation}
where $\sigma_{\epsilon}= 0.42 $ and $n_g = 39.3 $arcmin$^{-2}$ 
are the observed variance of galaxy ellipticities and number density,
respectively, yielding $\sigma_{\rm noise} = 0.020 \theta_G^{-1}$.

We repeated the randomization 100 times and compute the 
rms among them at each grid where $\kappa$ is computed.
Since $\kappa$ is Gaussian in its errors, this rms represents
1-$\sigma$ noise level on each grid, and thus the measured signal
divided by the rms gives the S/N ratio, $\nu$, at that grid. 
Thick contour in Fig.~\ref{kappa} shows the S/N-map with a 
smoothing scale of $\theta_G = 1'$. 
Threshold of the contour is set at 3.

One notes that numbers of dark halo candidates are detected.
For reference, number density of moderately bright galaxies are 
super-imposed as thin contour. It is surprising that correlation of
mass and light is clearly visible from this relatively simple analysis
on the single band image. A halo that has the highest significance is
designated (X) in Fig.~\ref{kappa}.
In this halo, the galaxy density also shows an excess as 
large as the other known clusters.
In fact, a spectroscopic observation of two member galaxies
implies that it is a cluster located at the redshift of 0.41. This
object is not listed on the NASA/IPAC Extragalactic Database (NED). 
Other follow-up observations are underway and detailed 
comparison of mass and light will be discussed in subsequent papers. 

\vspace{0.3cm}
\centerline{{\vbox{\epsfxsize=8.5cm\epsfbox{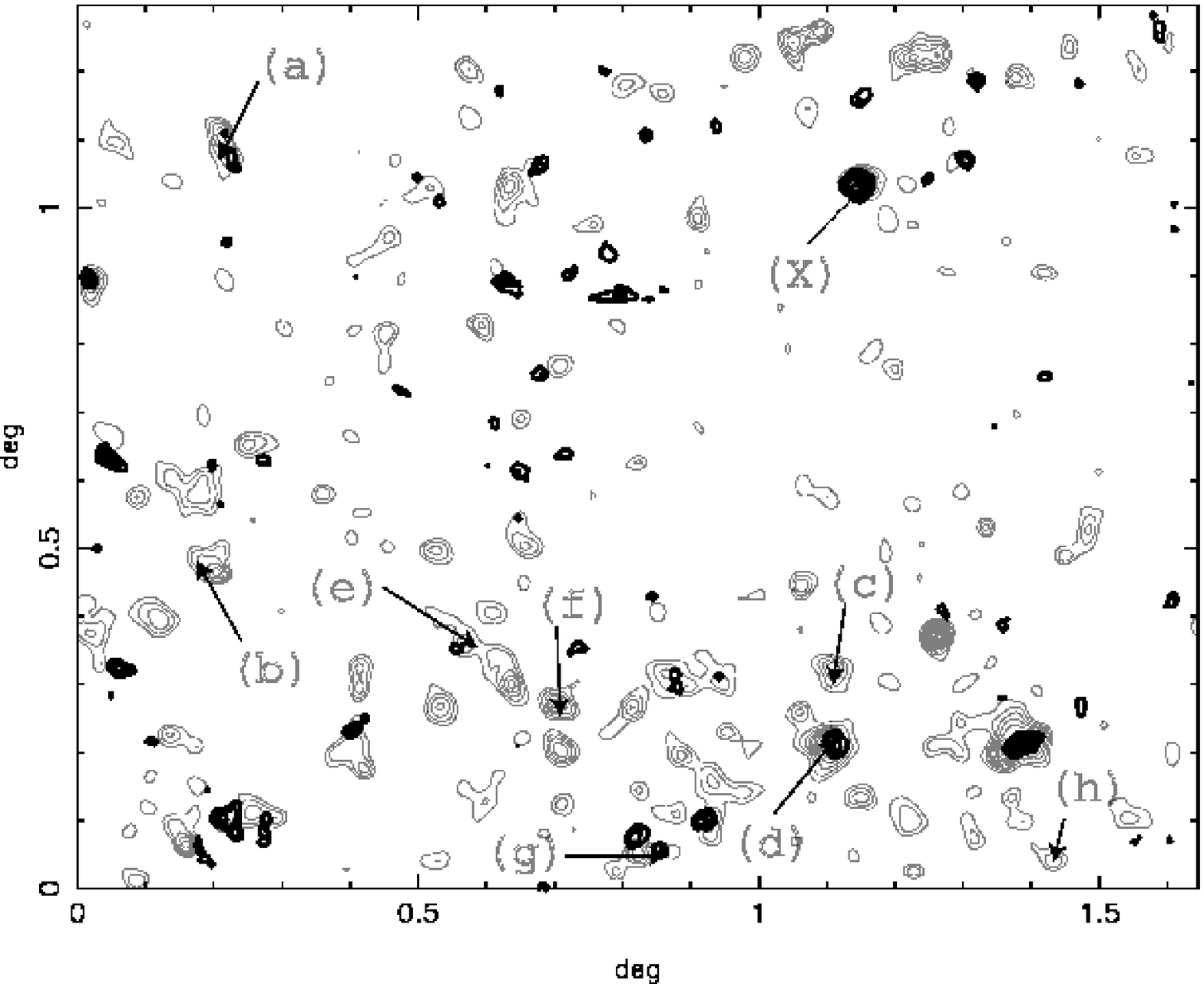}}}}
\figcaption{Thick contour represents the signal to noise ratio ($\nu$)
of convergence field. Data of $\nu > 3$ is shown. Thin contour shows
number density of moderately bright galaxies, $20 < R_{c} < 23$.
Symbols (a)-(h) are designated where clusters listed in the NASA/IPAC
Extragalactic Database (NED) are located and excess of galaxy number
density is visible in the map; (a)GHO1606+4346, (b)Abell2158 (0.13),
(c)GHO1601+4259 (0.54), (d)GHO1601+4253 (0.54), (e)GHO1604+4303,
(f)GHO1603+4256, (g)GHo1602+4245, (h)GHO1559+4242. The
redshift is shown in the bracket if known. Both maps are smoothed
with Gaussian kernel of $\theta_G = 1'$.
\label{kappa}}
\vspace{0.3cm}

Let us turn to statistical properties of the S/N-map, and we will 
compare the results with the model predictions. 
First, we examine the probability distribution function (PDF) of peaks.
Peaks are identified by pixels having a higher or lower value of $\nu$ 
than the all eight surrounding pixels as suggested by 
\cite{jainvanwaerbeke00}. 
The PDF of the peak heights is shown as thick histograms of left two
panels in Fig.~\ref{fmax_jvw_kappa}.
The thin histogram shows a noise PDF measured from $\kappa_{\rm noise}$ 
maps, the average of 100 randomizations is plotted and 
the error bars show the rms among them. The bimodal feature of 
the noise PDF is due to superposition of two contributions each 
of which is due to positive peaks and negative peaks (= troughs).
Excess counts over the noise PDF is apparent on both
positive ($\nu \ge 3$) and negative ($\nu \le -3$) side of the
histogram. The excess becomes more apparent as $\theta_G$ increases. 
The positive peaks imply the existence of dark matter halos 
whereas the natural explanation of the negative peaks seen is 
voids \citep{jainvanwaerbeke00}. This is the first observational 
evidence of the existence of voids in the matter distribution 
revealed by weak lensing.

Right panels of Fig.~\ref{fmax_jvw_kappa} show the peak
PDF measured from numerical experiments of the gravitational
lensing by large-scale structures.
The experiments were performed using the ray-tracing technique
combined with large N-body simulations and its details are described
in M\'{e}nard et al. (2002).
Briefly, $N$-body data from Very Large simulation
carried out by the Virgo Consortium (Jenkins et al 2001;
Yoshida, Sheth \& Diaferio 2001) were used to generate
the dark matter distribution.
A $\Lambda$CDM model ($\Omega_m=0.3$, $\Omega_\lambda=0.7$,
$h=0.7$ and $\sigma_8=0.9$) is assumed.
The multiple-lens plane ray-tracing algorithm was used to follow the
light rays  (Hamana \& Mellier 2001, see also Jain, Seljak \& White
2000 for technical detail). The lensing convergence and shear were 
computed for 1024$^2$ pixels with the pixel size of 0.25 arcmin for a 
single source plane of $z_s=1$.
We generated the simulated noisy $\kappa$ map following the procedure
described in \cite{jainvanwaerbeke00}.
The noise is modeled as a Gaussian random field with the variance
measured in the real data.
The peak is identified as the same manner as for the observed data.
Ten realizations of the numerical experiment were used to compute
the peak PDF (the total field size is $10 \times 4.27^2$
deg$^2$, much larger than the observed data, and thus the peak
distribution is very smooth).

\vspace{0.3cm}
\centerline{{\vbox{\epsfxsize=8.5cm\epsfbox{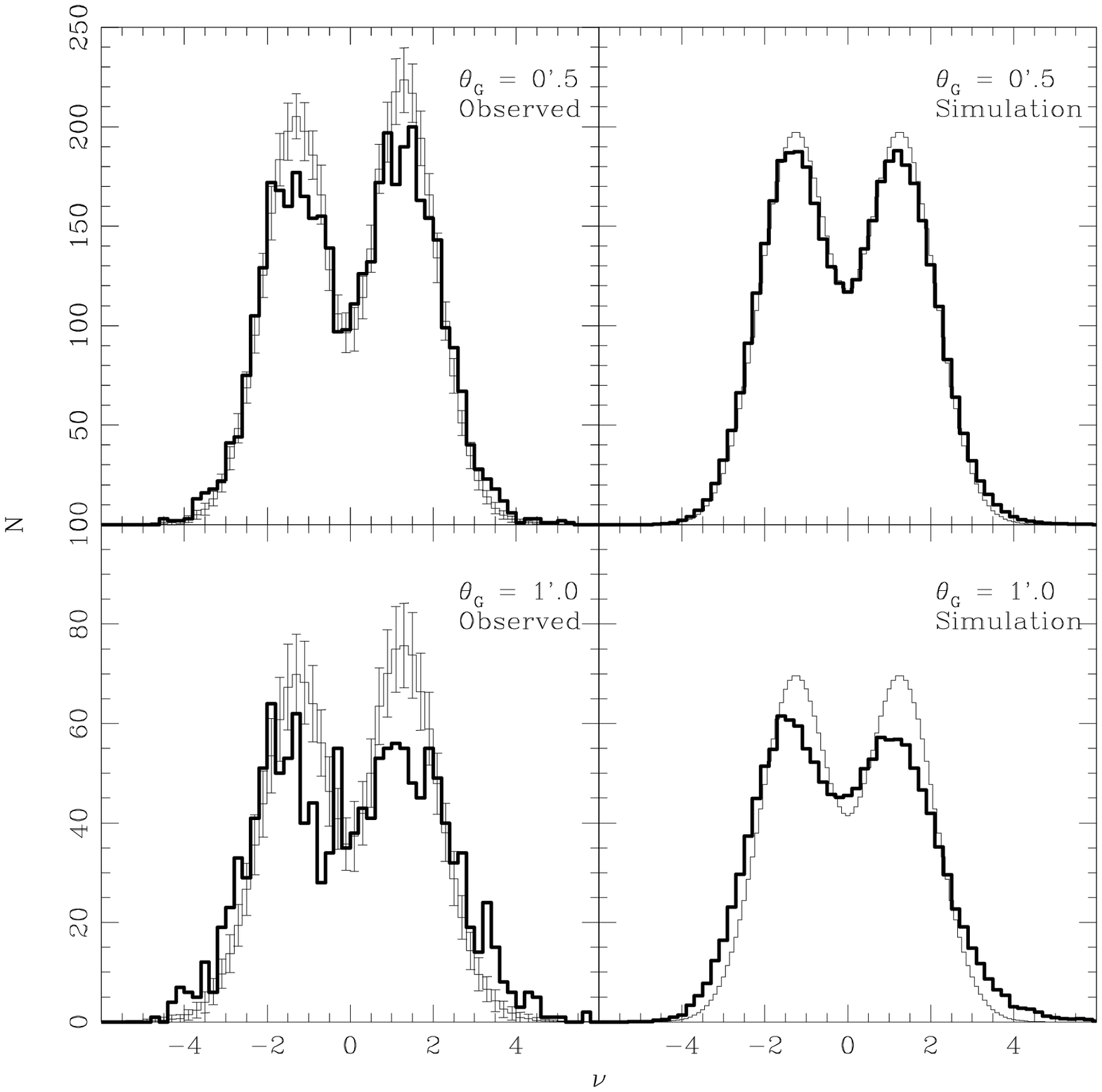}}}}
\figcaption{
Peak distribution function in $kappa$ map with smoothing scale
$\theta_G = 0'.5$ (top) and $1'$ (bottom). Thick solid histograms
on the left panels show the results obtained from the observed catalog
and thin solid histograms associated with error bars are averaged
results of 100 random realizations of galaxy orientation in the
catalog. The errors are rms of the randomization.
Results from $\Lambda$CDM model simulation are shown as thick
histograms on the right based on the observed parameters,
$\sigma_{noise} = 0.023\theta_G^{-1}$ assuming $z_s = 1.0$ and
their pure noise PDF is shown as thin
histograms.\label{fmax_jvw_kappa}}
\vspace{0.3cm}

The simulated PDF is very similar to the observed PDF.
Jain \& van Waerbeke (2000) pointed out that the peak PDF can be 
used to discriminate the cosmological model.
They demonstrated using the ray-tracing experiments of a high density 
($\Omega_m=1.0$, $\Omega_\lambda=0$, $\sigma_8 = 0.9$) and a low 
density($\Omega_m=0.3$, $\Omega_\lambda=0$, $\sigma_8 = 0.9$) models
that the high density model has broader distribution than the 
low density model, especially on the negative side.
It should be noticed that they adopted $\sigma_\epsilon=0.2$ which is 
about half of our measured value, therefore one should not directly 
compare our real PDFs with theirs (Figure 3 of \cite{jainvanwaerbeke00}).

In addition to the limited statistics of this work, theoretical 
prescription of the peak PDF is, however,  still premature
for placing constraint on the cosmological model in a quantitative 
manner. We therefore concentrate only on excess on the positive 
side here; the halo number count. We calculated number of the 
positive peaks, $N_{obs}$, in the observed S/N map exceeding 
$\nu_{\rm th}$. Halo number count is defined as 
$N_{halo} = N_{obs} - N_r \pm \sqrt{N_{obs} + \sigma_{N_r}^2}$ 
where $N_r$ and $\sigma_{N_r}$ is average peak counts obtained 
on the randomized catalogs and their variances, respectively.
The halo count is shown in Fig.~\ref{n_nu} as well as 
theoretical predictions which are constructed following 
Kruse \& Schneider (1999).
We assumed a high positive peak in $\kappa$ map comes from a 
single halo whose number density is described by Press-Schechter 
mass function  (Press \& Schechter 1974; we used its modified model 
by Sheth \& Tormen 1999).
We further assumed that all halos have a universal density profile
that Navarro et al. (1996) proposed (NFW); $\rho(r) = \rho_s
r^{-1}(1+r)^{-2}$. We also compared the results with another halo
profile, singular isothermal sphere model (SIS); 
$\rho(r) = \rho_s r^{-2}$. 
For the relation between the scale length of the halo and halo mass 
we adopted the fitting form by Bullock et al (2001).
The $\sigma_{\rm noise}$ measured from observed data 
is used to convert $\kappa$ to $\nu$.
Note that Kruse \& Schneider (1999) and Bartelmann et al. (2001) 
adopted $\sigma_\epsilon=0.2$, thus their predicted count is much larger 
than our prediction.

\vspace{0.3cm}
\centerline{{\vbox{\epsfxsize=8cm\epsfbox{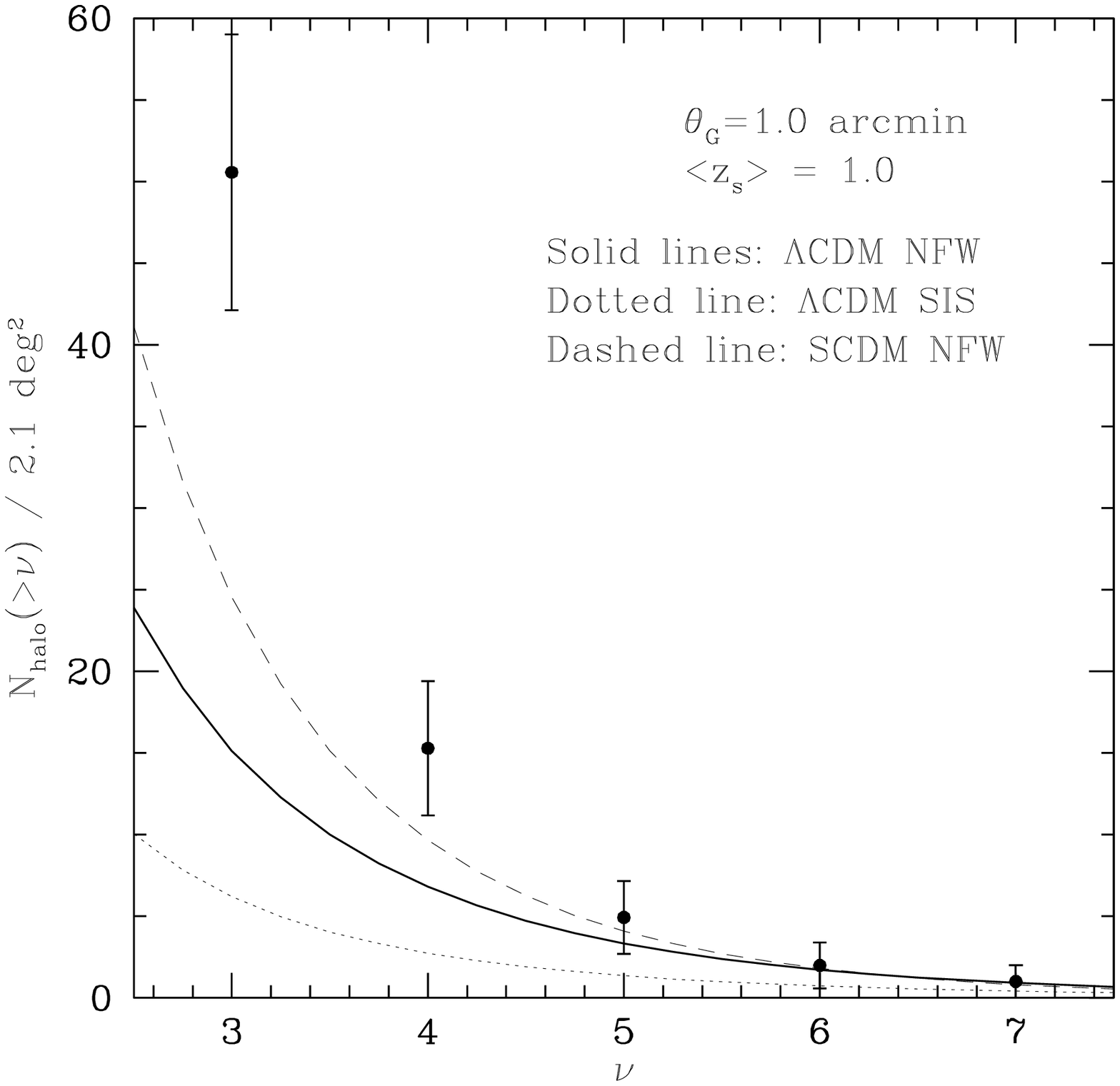}}}}
\figcaption{
Dark matter halo counts on the convergence field with $\theta_G =
1'$. The counts due to noise are estimated from the random
realizations and subtracted. Solid and dotted line show $\Lambda$CDM
model with the halo profile of NFW and SIS, respectively.
Dashed line presents SCDM with NFW profile. The model calculation
assumes a single source redshift of $z_s=1.0$. Broadening of the
redshift distribution makes the count slightly less but the effect is
a few percents at most as far as the empirical distribution
with reasonable parameters are assumed (e.g. Hamana et al. 2002).
\label{n_nu}}
\vspace{0.3cm}

First, if one considers only highly significant halos, say $\nu \ge
5$, the predicted halo number count does not depends strongly on the
background cosmology as is presented in Fig.3, where dashed line
and solid line shows the SCDM model ($\Omega_m=1$, $\Omega_\Lambda=0$,
$\sigma_8=0.6$, $h=0.5$) and $\Lambda$CDM ($\Omega_m=0.3$,
$\Omega_\Lambda=0.7$, $\sigma_8=0.9$, $h=0.7$), respectively.
The count, however, depends on the halo profile; the SIS predicts less
counts as is shown in the dotted line in the figure. The observed
count prefers NFW profile although the poisson error in $\nu \ge 6$ is
large in this work. It is demonstrated here that number count of
highly significant halos can be an useful probe for the dark matter
halo profile because of the independence of background cosmology.
Meanwhile, the dependence becomes clearly visible in the counts of 
moderately significant haloes ($\nu < 5$); the SCDM predicts more 
counts than LCDM does due mainly to larger $\Omega_m$. Our observed 
result shows high number counts which favors high density universe.
This result, however, cannot be taken at face value since the
large excess is apparent in Fig.\ref{fmax_jvw_kappa} as a spike
around $\nu \sim 3.5$. Therefore, the excess count in the moderately
significant halos is likely due to the cosmic variance. Comparison
of data taken on multiple (say 5 to 10) uncorrelated fields of
this size is necessary to evaluate the contribution from the cosmic
variance and eventually to discuss cosmological models.

In summary, the present work clearly demonstrates that weak lensing
technique is indeed a promising way to probe matter distribution
purely through the mass. We detected not only the high positive 
peaks due to massive halos but also the excess in the peak PDF on 
the negative side which is the first observational evidence of the 
voids in dark matter distribution. Dark matter halo counts is proved
to be the useful tool to test the paradigm of the structure formation
and cosmological models. 
The cosmic variance that we encountered in this work should be 
evaluated by the data obtained by future high resolution wider field 
imaging survey.

\acknowledgments
We are grateful to L. van Waerbeke for helpful discussions about 
galaxy shape analysis, and M. Bartelmann for useful discussions.
We thank Richard Ellis and Alexandre Refregier for their careful
reading of the manuscript and the comments. T.H. F.N. and M.O. 
acknowledge support from Research Fellowships of the Japan Society 
for the Promotion of Science.

\end{document}